\documentclass[runningheads]{llncs}
\usepackage{graphicx}
\usepackage{amssymb}
\usepackage{amsmath}
\usepackage{cite}
\usepackage{array}
\usepackage{float}

\begin{document}

\title{Extracting Relations Between Sectors}

\author{Atakan Kara \inst{1} \and
F. Serhan Dani\c{s}\inst{2} \and
G\"{u}nce Keziban Orman \inst{2} \and
Sultan Nezihe Turhan \inst{2}}
\authorrunning{A. Kara \textit{et al.}}
%
\institute{Kariyer.net, Ar-Ge, Istanbul, Turkey\\
\email{atakan.kara@kariyer.net} \and
Department of Computer Engineering, Istanbul Turkey\\
\email{\{sdanis,korman,sturhan\}@gsu.edu.tr}}
\maketitle              
\begin{abstract}
The term "sector" in professional business life is a vague concept since companies tend to identify themselves as operating in multiple sectors simultaneously. This ambiguity poses problems in recommending jobs to job seekers or finding suitable candidates for open positions. The latter holds significant importance when available candidates in a specific sector are also scarce; hence, finding candidates from similar sectors becomes crucial. This work focuses on discovering possible sector similarities through relational analysis. We employ several algorithms from the frequent pattern mining and collaborative filtering domains, namely negFIN, Alternating Least Squares, Bilateral Variational Autoencoder, and Collaborative Filtering based on Pearson's Correlation, Kendall and Spearman's Rank Correlation coefficients. The algorithms are compared on a real-world dataset supplied by a major recruitment company, Kariyer.net, from Turkey. The insights and methods gained through this work are expected to increase the efficiency and accuracy of various methods, such as recommending jobs or finding suitable candidates for open positions.
\end{abstract}
\section{Introduction}
Businesses are classified into sectors based on the part of the economy they operate in. Whereas the same job titles are used for positions in different sectors, their requirements, responsibilities, and objectives may differ significantly. Therefore, people who work in a sector for long periods gain experience based not only on their specific position but also on the sector-specific dynamics, making them more valuable hires in that sector. So, it is crucial to consider sectors while working on various problems in the human resources (HR) domain, such as recommending jobs to candidates or finding the right fit to fill a vacancy. Moreover, it is essential to consider similar sectors when candidates working in a particular sector are scarce. Such vacancies can be filled by candidates working in different sectors with similar characteristics.

Accordingly, we investigate finding relationships between sectors using the data obtained from Kariyer.net\footnote[1]{Kariyer.net is a recruitment platform in Turkey.}. We aim to extract similarity relations among sectors that have similar dynamics. Outputs of this work can be used as additional features for some of the machine learning problems in the HR domain, such as job recommendation \cite{job_rec_1, job_rec_2, job_rec_3}, candidate searching \cite{candidate_ranking_1, candidate_ranking_2, candidate_ranking_3},  estimating the duration of employment \cite{job_duration}, and predicting employee turnovers \cite{turnover_1, turnover_2}. 

While finding similarities between entities is a hot topic in the HR domain, research on sector-based similarity extraction or inference is limited. For example, the similarity of jobs is studied in the work of Encheva \cite{rank_job_appl} to rank job applications based on the similarity of qualified candidates who did not accept the offer. In addition, Dave \textit{et al.} study extracting both job similarities and skill similarities to recommend jobs using representation learning models trained on job-transition, job-skill, and skill co-occurrence graphs \cite{combined_job_skill}.

This work compares various algorithms from the Frequent Itemset Mining (FIM) and Collaborative Filtering (CF) literature to extract similar sector relations. FIM is an essential data mining method to extract patterns and frequently occurring items. Patterns and knowledge extracted by the FIM algorithms significantly benefit decision-making processes \cite{freq_itemset_mining}. A typical application of this method is finding the set of products commonly purchased together in a supermarket. Based on the assumption that frequently co-occurring sectors in the profiles of the companies have similar dynamics, we apply an FIM algorithm to find sector pair patterns that frequently occur.

CF is a method used for recommending items by utilizing the collaborative behaviors of the users. CF algorithms can be grouped into three subcategories: user-based, item-based, and combination of both user- and item-based methods. User-based Collaborative Filtering (UBCF) algorithms are widely used for recommending items to customers based on the ratings of similar users. They predict the ratings of unrated items according to the ratings available from other users. Ratings from similar users to the target user contribute more to predicting the final rating. The main steps of UBCF are selecting neighbors, predicting ratings, and making recommendations. The neighbor selection step includes finding the similarity among all users to the target user. Afterward, ratings made by users with the highest similarity scores are extracted. This data is used to predict the rating score of each item that the target user has not rated yet. Finally, items with the highest predicted ratings are recommended to the user.

Similar to the UBCF, item-based collaborative filtering (IBCF) is a popular family of algorithms for recommending items to users. However, unlike UBCF, instead of recommending based on the ratings of similar users, IBCF focuses on item similarities and predicts ratings based on ratings of similar items. Ratings of similar items contribute more to the predicted rating of the target item. First, it computes the similarity of items based on the ratings provided by the target user and then selects the $K$ items with the highest similarity scores. Ratings of the selected $K$ items are then used to predict the rating scores for items the target user has not yet rated. Lastly, the items with the highest predicted ratings are recommended.

User-based and item-based collaborative filtering algorithms suffer from two main problems: data sparsity and scalability. Combinations of user-and item-based algorithms are proposed to address these challenges and have been shown to give better recommendations \cite{hybrid_user_item_cf}. They predict the rating of each item based on ratings given by the target user and other users. However, instead of focusing on similar users or similar items, they utilize both.

None of the algorithms mentioned above can be directly applied to our dataset for extracting sector similarities. As a contribution, we present reductions we made to transform the extracting similar sector relations problem into the problems of the well-studied algorithms presented in this paper. Moreover, we compare and discuss the results of these algorithms on a real dataset as a side contribution.

In the rest of the paper, Section~\ref{sec:methodology} describes the algorithms and their transformation to match our needs. Then, we present the dataset used for training and evaluating the algorithms, the details of the experiments, and the experimental results in Section~\ref{sec:experiments}. Finally, we conclude with the insights gained and future directions in Section~\ref{sec:conclusion}.

\section{Methodology}
\label{sec:methodology}

In this section, we first define the similar sector notion and then present the algorithms used to extract these relations conveniently. We will denote the individual sectors with $s_i \in \mathbb{S}$, where  $\mathbb{S}$ is the set of all sectors. The relation between two sectors, $s_i$ and $s_j$, will be defined by $\mathcal{R}_{ij} \in \{0,1\}$, which indicates whether $s_i$ is related to $s_j$. Since the strengths of the relations will not be considered, they are represented as binary. Our goal is to find all the $\mathcal{R}_{ij}$ in $\mathbb{S}$ from the common attributes of the sectors.



Even though the ``similar sector'' concept is not well defined, human experts are consistent and good at distinguishing similar sectors from others. For example, humans relate the ``Building Sector'' to the ``Infrastructure Sector''. Based on the domain-specific knowledge gained from the human experts, we conclude that the sector $s_i$ is similar to the sector $s_j$, corresponding to $\mathcal{R}_{ij} = 1$, if the two sectors have similar internal dynamics or share required sector-specific skills, or an experienced individual in one sector becomes a valuable candidate in the other.

In the rest of the section, we present an FIM algorithm and five different algorithms from the (CF) domain. Unlike FIM, CF algorithms do not directly extract relations between items. Instead, they either implicitly use similarities among items or do it as an intermediate step. We transform our dataset and reduce our problem to a CF problem, which is explained in detail in the following sections. This transformation enables us to benefit from the rich CF literature. We share the reductions used to transform our problem into the problems of well-studied algorithms.

\subsection{Frequent Itemset Mining (FIM)}
\label{sec:FIM}
We define a set $\mathcal{S} = \{s_1, \cdots, s_n\}$ of \textit{items} that represent the sectors and a database $\mathcal{T} = (t_1, \cdots, t_n)$ of records for set of sectors companies are operating in. For example, an item may be the Retail sector, and a record $t_i$ might consist of the Retail, E-Commerce, and Clothing Sectors. First, any frequent itemset given by the FIM algorithm that does not have exactly two items is dropped. Then, the support values extracted with the FIM algorithm for each pair are assigned as the "similarity score" of that sector-sector pair. Pairs without support values are assigned a similarity score of zero, which corresponds to $\mathcal{R}_{ij} = 0$. Then, for each sector $s_i$, we pick the top $K$ sectors with the highest support values and mark them as similar and the rest as dissimilar. As a result, we assign the  $\mathcal{R}$ relationship between $s_i$ and $s_j$ if $s_j$ is one of the top $K$ sectors with the highest similarity score to $s_i$. Lastly, we assign $\mathcal{R}_{ij} = 0$ for the rest of the pairs for each sector.

\paragraph{negFIN Algorithm} proposed by Aryabarzan~ \textit{et al.} \cite{negFIN} is used to find all frequent itemsets. It forms the search space with the help of a  set-enumeration tree. First, it generates a bitmap code of the itemsets-tree identifying 1-itemsets with their Nodesets and creates the first level of the set-enumeration tree. Each node in the set-enumeration tree has an item-name and children-list field. Additionally, the Itemset of each node is defined as the itemsets of the father node and the item-name of the current node. Then, it identifies frequent 2-itemsets with their NegNodesets and constructs the second level of the set-enumeration tree. Afterward, the frequent itemsets with support values greater than two are identified with their NegNodesets, and the rest of the set-enumeration tree is created. Finally, it uses a promotion method to prune the frequent itemsets in the set-enumeration tree. As a result, the negFIN algorithm achieves itemset support calculation in $O(n)$. Since negFIN is the fastest algorithm \cite{negFIN}, we decided to use it to extract similarity relations from our dataset.

\subsection{User-based Collaborative Filtering (UBCF)}
\label{sec:UBCF}
UBCF algorithms model similarities between users to make recommendations. To benefit from them, we reframe our problem to recommend sectors to companies operating in a single sector. So, companies are treated as users, and sectors are treated as items. Here, for $M$ sectors and $N$ companies, we define a matrix $\bf Y \in  \mathbb{R}^{M \times N}$ representing the sectors to which the companies belong. Each entry $y_{m,n} = r$  in the matrix indicates whether company \textit{n} is working on sector \textit{m}, where $\textit{r} \in \{1, 0\}$. First, synthetically generated companies operating in a single sector are added to the dataset. So, for each sector, we ensure to have an isolated company that only operates in that sector. After running the UBCF algorithm for training, we get the prediction scores on whether a sector should be recommended to that company for each synthetically created company. Assuming that synthetically created company $C_i$ is operating in the sector $s_i$,  we assign relation $R$ between sector $s_i$ and $s_j$ if sector $s_j$ is in the top $K$ recommendations for the synthetic company $C_i$.

\paragraph{Alternating Least Squares (ALS)} is a matrix factorization algorithm widely used for collaborative filtering \cite{ALS}. Given an $m \times n$ matrix $\mathrm{R}$ representing ratings made by users, ALS algorithm factors $\mathrm{R}$ into two smaller matrices $\mathrm{U}$ of size $m \times l$ and $\mathrm{P}$ of size $l \times n$. Then, the rating of item $j$ given by user $i$ is predicted as
\begin{equation}
r_{u i} \approx u_i \cdot p_j^T 
\end{equation}

\subsection{Item-based Collaborative Filtering (IBCF)}
IBCF algorithms aim to model item similarities based on the interactions of the users. We feed $\bf Y \in  \mathbb{R}^{M \times N}$ matrix defined in Section~\ref{sec:UBCF} to the IBCF algorithms, which causes them to calculate sector similarities implicitly. Normally, an IBCF algorithm would create recommendations for a user based on similar items with which the user has interacted. Since we are interested in extracting similar sector relations, we omit the recommendation step and save the item similarities matrix created by the IBCF algorithms. Then, for each sector $s_i$, we sort the corresponding column in the item similarities matrix, pick top $K$ sectors, and assign relation $\mathcal{R}_{ij} = 1$ if $s_j$ is within top the $K$ sectors.

\paragraph{Pearson Correlation Coefficient (Pearsons's $r$)} is a measure to assess the linear correlation between itemsets. The coefficient is determined by calculating the ratings of two items and normalizing them over their standard deviations. We formulate as follows:
\begin{equation}
sim(i, j) = \frac{{}\sum_{i \in com(i, j)} (r_{u,i}- \overline{r_u})(r_{u,j} - \overline{r_u})}
{\sqrt{\sum_{i \in com(i, j)} (r_{u,i}- \overline{r_u})^2 } \sqrt{\sum_{i \in com(i, j)} (r_{u,j}- \overline{r_u})^2 }}
\end{equation}
where $sim(i, j)$ is the similarity between items $i$ and $j$, $com(i, j)$ is the set of users that rated both item $i$ and $j$, $r_{u,i}$ is the rating of the $i$\textsuperscript{th} element rated by user u, and $\overline{r_u}$ is the average rating of user $u$.

\paragraph{Kendall Rank Correlation Coefficient (Kendall's $\tau$)} is a measure used to evaluate the ordinal association between two itemsets. It is a ranking-oriented measure where the similarity between two items is measured with the ordering of ratings given by two users. The coefficient is calculated as follows:
\begin{equation}
\tau = \frac{C - D}{\binom{n}{2}}
\end{equation}
where $C$ is the number of concordant pairs and $D$ is the number of discordant pairs defined as:

\[C = \mid\left \{(i, j) \mid x_{i}<x_{j} \text { and } y_{i}<y_{j}\right\}\mid\]
\[D = \mid\left\{(i, j) \mid x_{i}<x_{j} \text { and } y_{i}>y_{j}\right\}\mid\] 

\paragraph{Spearman's Rank Correlation Coefficient (Spearman's $\rho$)} is a nonparametric measure to evaluate the rank correlation between itemsets. It evaluates how well the relationship between itemsets can be represented with a monotonic function. It can be calculated as
\begin{equation}
\rho = 1- {\frac {6 \sum d_i^2}{n(n^2 - 1)}}
\end{equation}
where $d_i$ is the difference between two ranks of each rating and n is the number of ratings.

\subsection{Combination of User-based and Item-based Collaborative Filtering}
We applied the same strategy for reducing user-based collaborative filtering algorithms to our problem, as explained in Section~\ref{sec:UBCF}. We reformulate the problem to recommend sectors to companies operating in a single sector. We add synthetically generated companies operating in a single sector to the dataset. Then, we run the algorithm for each synthetically created company $C_i$ operating in the sector $s_i$ and assign relation $R$ between sector $s_i$ and $s_j$ if sector $s_j$ is in the top $K$ recommendations.

\paragraph{Bilateral Variational Autoencoder for Collaborative Filtering (BiVAE)}
is a CF algorithm built upon a generative model of user-item interactions and corresponding user and item inference models \cite{BiVAE}. Both generative and inference models are parametrized by multilayer perceptrons (MLP). The generative model draws latent variables from standard multivariate isotropic Gaussian priors. Inference and learning are approximated by Evidence Lower Bound (ELBO). BiVAE is optimized by employing a reparametrization step \cite{reparam_trick_1, reparam_trick_2} and building an unbiased Monte Carlo estimator \cite{BiVAE}.

\section{Experimental Setup and Results}
\label{sec:experiments}

\subsection{Dataset}
\label{sec:Dataset}

The companies that want to post job announcements create profiles on the web portal of Kariyer.net. Alongside their business-related information, they specify the sectors they are operating in by picking one or multiple items from a predetermined list of sectors accumulated at Kariyer.net over the years. Therefore, the resulting dataset consists of the unique identities of each company with a list of sector identities. A sample dataset view is shown in \tablename~\ref{dataset_table}. We also provide the statistics of this dataset in \tablename~\ref{dataset_attributes}.  

\begin{table}
\caption{A sample from the dataset.}
\label{dataset_table}
\centering
\begin{tabular}{|l|l|l|} 
\hline
\multicolumn{1}{|c|}{Company ID} & \multicolumn{1}{c|}{Sector IDs}                                                   & \multicolumn{1}{c|}{Sector Names}                                                                                        \\ 
\hline
1718612                          & 030000000,029003000                                                               & Printing - Publishing, Cinema - Theatre                                                                                  \\ 
\hline
1718051                          & 20009000                                                                          & Building                                                                                                                 \\ 
\hline
1717919                          & 003000000                                                                         & Electric \& Electronic                                                                                                     \\ 
\hline
1719337                          & 040001000,020011000                                                               & Waste Management, Treatment                                                                                              \\ 
\hline
1718896                          & \begin{tabular}[c]{@{}l@{}}001000000,001005000,\\023003000,026007000\end{tabular} & \begin{tabular}[c]{@{}l@{}}Informatics, Software, Finance and Investment \\Consultancy,~Financial Services\end{tabular}  \\ 
\hline
1719789                          & 018000000                                                                         & Pharmaceuticals Industry                                                                                                 \\ 
\hline
1737591                          & 019000000                                                                         & Advertising                                                                                                              \\
\hline
\end{tabular}
\end{table}

\begin{table}
\caption{Attributes of the dataset.}
\label{dataset_attributes}
\centering
\begin{tabular}{|l|l|} 
\hline
Number of Companies       & 79884   \\ 
\hline
Number of Sectors         & 394     \\ 
\hline
Average Number of Sectors & 1.544  \\ 
\hline
SD of Number of Sectors   & 1.211  \\
\hline
\end{tabular}
\end{table}

\begin{figure*}
\centerline{\includegraphics[width=\textwidth]{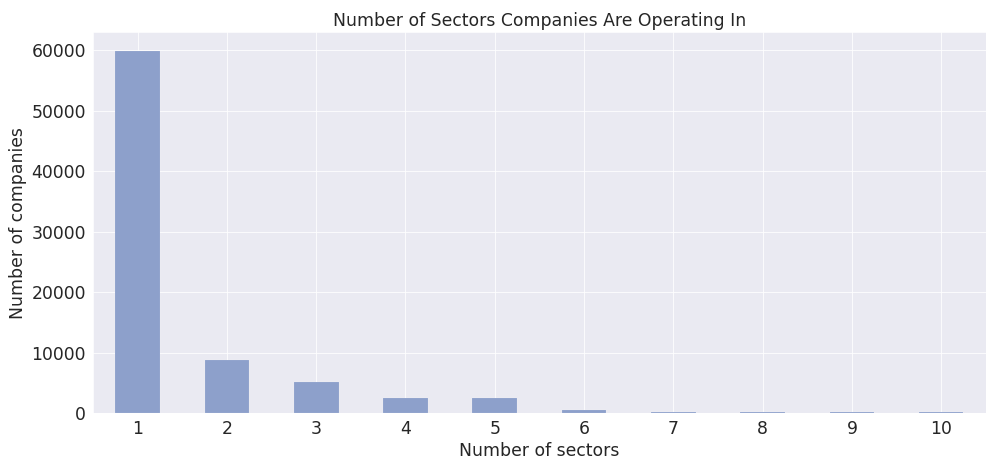}}
\caption{Number of sectors companies are operating in.}
\label{number_of_sectors}
\end{figure*}

As shown in Figure \ref{number_of_sectors}, most companies operate in a single sector. Since all of the algorithms studied in this work rely on extracting similarity relations from co-occurrences and correlations, those companies do not contribute to the results. The rest of the companies mostly operate in between 2-5 sectors. Since both Collaborative Filtering and Frequent Itemset Mining algorithms suffer from data sparsity, the distribution of the number of sectors makes the extracting sector relations task more challenging. 

\subsection{Experiments}

To evaluate the results, we labeled the sectors that have similar relations. Since companies add sectors to their profiles without explicitly pointing out the sector relations, the labeling step was required to create a ground truth. There exist 394 unique sectors in the database, which makes 155,236 many possible relations. So, labeling all possible pairs is an impractical and time-consuming task. Instead of labeling every possible combination, we combine the models' predictions and assign a score to possible similarity relations according to the output of the models. We then pick the sector pairs with scores that exceed the threshold. This approach enables us to consider only a small partition of the dataset. Finally, we ask experts to assess whether the two sectors are similar and label each sector pair by the vote of the majority of the experts. We assume that this approach represents the ground truth relations, which enables us to evaluate the models.

The models' performances are evaluated using Precision@$K$, Mean Reciprocal Rank (MRR), and Mean Average Precision at $K$ (MAP@$K$) metrics. We binarize the results of collaborative filtering algorithms by assigning scores of 1 to the top 10 results and 0 to the rest. This technique is applicable in our case since we will consider only the first ten similar sectors when working on downstream tasks. 

For the negFIN algorithm, we used java implementation of the algorithm in the SPMF Library \cite{spmf}. For the BiVAE and ALS, Microsoft Recommenders Library \cite{microsoft_rec} was used.

\subsection{Results}
We applied negFIN, ALS, BiVAE and Pearson, Kendall, and Spearman correlation-based CF algorithms to find similarity relations. From the output of the negFIN algorithm, we picked frequent itemsets consisting of only two sectors. The first ten itemsets with the highest support values can be seen in Table~\ref{negfin_top_10}.

\begin{table}
\caption{The firtst 10 itemsets consisted of two sectors with the highest support values given by the negFIN algorithm.}\label{negfin_top_10}
\centering
\begin{tabular}{|l|l|c|} 
\hline
Sector\#1                          & Sector\#2                           & \multicolumn{1}{l|}{Support}  \\ 
\hline
Merchandising                      & Retail                              & 6256                          \\ 
\hline
Clothing                           & Textile                             & 5679                          \\ 
\hline
Hotel and Accommodation~Management & Tourism                             & 5639                          \\ 
\hline
Ready-to-wear                      & Textile                             & 5014                          \\ 
\hline
Software                           & Informatics                         & 4146                          \\ 
\hline
Retail                             & Textile                             & 3418                          \\ 
\hline
Automotive                         & Manufacturing / Industrial Products & 2292                          \\ 
\hline
Automotive Aftermarket             & Automative                          & 2213                          \\ 
\hline
IT Machinery and
  Equipment       & Manufacturing / Industrial Products & 2103                          \\ 
\hline
Fast-moving consumer goods         & Food                                & 1736                          \\
\hline
\end{tabular}
\end{table}

At first glance, this table supports the similar sector relation described in Section~\ref{sec:methodology}. For example, Merchandising-Retail, Clothing-Textile, Hotel and Accommodation Management-Tourism sector pairs have similar internal dynamics and experience gained from one would make a candidate more valuable in the other.  
Effects of data sparsity discussed in Section~\ref{sec:Dataset} are observed in the correlations. Since most of the companies are operating in a single or few numbers of sectors, there are not any strong correlations exist. Moreover, many sectors do not have common companies operating in both sectors.

\begin{figure}[hbt!]
\centerline{\includegraphics[scale=0.3]{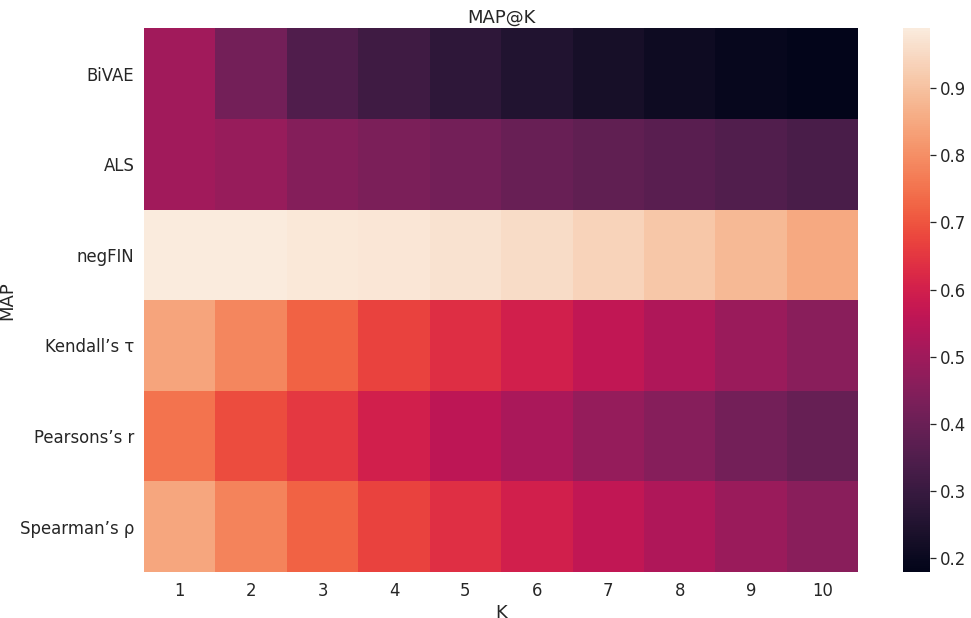}}
\caption{Heatmap of the MAP@$K$ scores of the algorithms.\label{map_heatmap}}
\end{figure}

\begin{figure*}
\centerline{\includegraphics[scale=0.3]{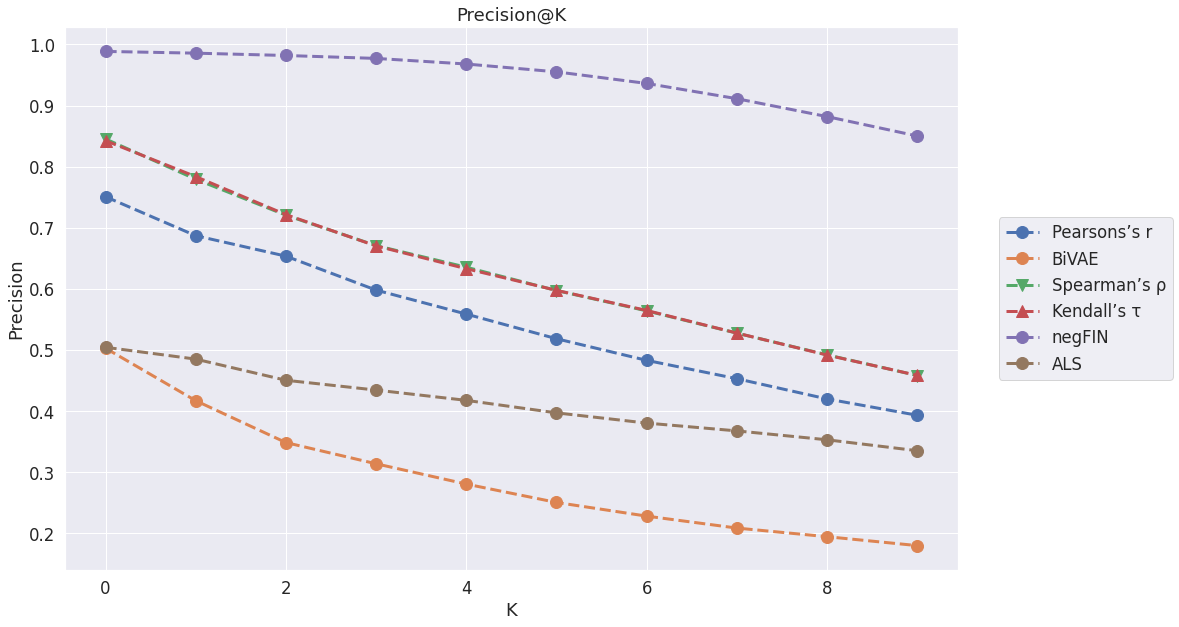}}
\caption{Precision@$K$ scores of the algorithms.}
\label{precision_plot}
\end{figure*}

Even though the negFIN algorithm is not intended for this task, with proper tuning, as explained in Section~\ref{sec:FIM}, it can be used to find the most related items from a given set. On the other hand, user-based and hybrid collaborative filtering algorithms, ALS and BiVAE, do not give promising results. Since those algorithms offer recommendations based on the extracted similarities between companies, low prediction and MAP results might be explained by the fact that those models try to find user-item similarities explicitly and extract item-item similarities from them. The negFIN algorithm's success indicates that the FIM algorithms' underlying assumption holds to a certain extent in the studied dataset. That is, frequently co-occurred sectors in the database $\mathcal{T}$ are related. Even though Collaborative Filtering algorithms could also find the relationships between co-occurred items, inductive bias in the negFIN algorithm caused a better estimation of relations in the dataset.

Item-based collaborative filtering algorithms perform better than User-based and Hybrid Collaborative Filtering approaches. This is an expected outcome since they explicitly model item-item relations to make recommendations instead of implicitly inferring from user-item relations. However, they do not attain the performance of the Frequent Itemset Mining approach, which indicates that they are not as good as the negFIN algorithm at finding item-item relations from companies' set-value typed sector data. We also observe that MAP@$K$ scores of the algorithms decrease as the value of $K$ increases. Since finding similar sectors from a limited set gets more difficult as $K$ increases, this result was expected. The MAP@$K$, Precision@$K$ and MRR results for $K=5$ and $10$ are given in Table \ref{results_table} .

\begin{table}[hbt!]

\caption{Scores of the algorithms evaluated on Precision@$K$, MAP@$K$ and MRR.}\label{results_table}
\centering
\begin{tabular}{|p{0.18\textwidth}|p{0.15\textwidth}|p{0.15\textwidth}|p{0.15\textwidth}|p{0.15\textwidth}|p{0.15\textwidth}|} 

\cline{2-6}
\multicolumn{1}{l|}{} & P@5    & P@10   & MAP@5  & MAP@10  & MRR     \\ 
\hline
Pearson's $r$              & 0.609 & 0.463 & 0.559 & 0.393 & 0.827  \\ 
\hline
Spearman's $\rho$             & 0.681 & 0.522 & 0.635 & 0.458  & 0.889  \\ 
\hline
Kendall's $\tau$               & 0.678 & 0.522 & 0.522 & 0.458  & 0.888  \\ 
\hline
negFIN                & \textbf{0.968} & \textbf{0.850} & \textbf{0.968} & \textbf{0.850}  & \textbf{0.989}  \\ 
\hline
ALS                   & 0.492 & 0.431 & 0.280 & 0.335  & 0.640  \\ 
\hline
BiVAE                 & 0.350 & 0.256 & 0.280 & 0.180  & 0.607  \\
\hline
\end{tabular}
\end{table}

Kendall and Spearman's Rank Correlation Coefficient had highly similar results on MAP@$K$, Precision@$K$ and MRR metrics which can be seen from Table \ref{results_table}. Colwell \textit{et al.} \cite{spearman_vs_kendall} have reported that Kendall and Spearman's Rank Correlation Coefficient usually produce highly similar results. Therefore, getting close numbers from the evaluation metrics we used was expected. According to all studied performance metrics shown in Table~\ref{results_table}, negFIN outperforms other algorithms up to 50\%. The lowest performance is observed in BiVAE. Correlation-based IBCF algorithms performed similarly to each other. Accordingly, discovering frequent patterns of sectors from companies' declared sectors list data set is a good starting point for discovering sector similarities. While correlation analysis is naturally developed for finding possible relations between two variables or collaborative filtering is also dedicated to discovering possible collaboration between items, they seem to be one step behind compared to FIM methodology for our problem set.

\section{Conclusion}
\label{sec:conclusion}
In this work, we study the problem of extracting sector relationships from a real-world dataset obtained from a major recruitment platform. We applied several CF algorithms that use well-known correlation coefficients, such as Pearson's $r$, Kendall's $\tau$ and Spearman's $\rho$, along with negFIN, BiVAE and ALS. Because the CF algorithms are not directly applicable to our problem, the inputs should be transformed accordingly, after which the sector similarity task is reduced to a CF task. According to the results, the negFIN algorithm outperforms other methods and yields the best set of sector relationships on the dataset.

In future work, the findings can be applied to the other subtasks in the HR and recruitment domain, such as job recommendation, candidate searching, estimating the duration of employment, and predicting employee turnover. Moreover, this work can be enriched by assigning sector similarities instead of binary relationships.

\bibliography{references}
\bibliographystyle{splncs04}

\end{document}